\documentclass[english,A4]{IEEEtran}
\usepackage[T1]{fontenc}
\usepackage[latin9]{inputenc}
\usepackage{amsthm}
\usepackage{amstext}
\usepackage{amssymb}
\usepackage{graphicx}
\usepackage{enumerate}
\usepackage{mathtools}
\usepackage{physics}
\makeatletter
\theoremstyle{definition}
\newtheorem{thm}{\protect Theorem}
\theoremstyle{definition}
\newtheorem{lmm}{\protect Lemma}
\theoremstyle{definition}
\newtheorem{defn}{\protect Definition}
\theoremstyle{definition}

\makeatother

\usepackage{babel}

\begin{document}

\title{Robust Gaussian Joint Source-Channel Coding Under the Near-Zero Bandwidth Regime}

\author{Mohammadamin Baniasadi and Ertem Tuncel\\
University of California, Riverside, CA. Email: mohammadamin.baniasadi@email.ucr.edu, 
ertem.tuncel@ucr.edu\vspace*{-0.8cm}
}
\maketitle
\thispagestyle{empty}

\begin{abstract}
Minimum power required to achieve a distortion-noise profile, i.e., a function indicating the maximum allowed distortion value for each noise level, is studied for the transmission of Gaussian sources over Gaussian channels under a regime of bandwidth approaching zero. 
A simple but instrumental lower bound to the minimum required power for a given profile is presented. For an upper bound, a dirty-paper based coding scheme is proposed and its power-distortion tradeoff is analyzed. Finally, upper and lower bounds to the minimum power is analyzed and compared for specific distortion-noise profiles, namely rational profiles with order one and two.

\end{abstract}

\vspace{-3mm}
\section{Introduction}

We consider the classical scenario of lossy transmission of a Gaussian source over an additive white Gaussian noise (AWGN) channel, where the channel input constraint is on power.
We also do not assume any feedback in our system model.
When the channel noise variance $N$ is fixed, it is well-known (thanks to the famous separation theorem) that the minimum distortion that can be achieved with input power $P$ is given by
\begin{equation}
\label{dformula}
D_\kappa = \frac{1}{\left(1+\frac{P}{N}\right)^\kappa}
\end{equation}
where $\kappa$ is the bandwidth factor (with a unit of {\em channel uses per source symbol}). 

In this work, however, we instead consider the scenario where $N$ is not known at the transmitter (but known at the receiver as usual) and can assume any positive value $N>0$.
The system is to be designed to combat the unknown level of noise and comply with a {\em distortion-noise profile} ${\cal D}_\kappa(N)$, i.e., achieve 
\[
D_\kappa(N) \leq {\cal D}_\kappa(N) 
\]
for all $N>0$, while minimizing its power use, where $D_\kappa(N)$ denotes the achieved distortion at noise level $N$.
This setting reflects a very adverse situation in which even though the channel may be originally of very high quality ($N\approx 0$), it could be suffering from occasional interferences of a wide spectrum of noise levels (including $N\gg 0$). 

This scenario was previously tackled in~\cite{ermanenergy} in the context of infinite bandwidth, i.e., $\kappa\rightarrow\infty$, where {\em energy} naturally replaces power as the currency (see \cite{Jain2012}, \cite{Jiang2014}, \cite{Koken2016}, \cite{Koken2017}, and \cite{AminArxiv} for other work on energy-distortion tradeoff). Here, we address the other extreme, where the bandwidth is severely limited, i.e., $\kappa\approx 0$.  
This near-zero bandwidth condition might arise in cases where too many devices (e.g., in Internet-of-Things networks) share the same communication medium through multiplexing (e.g., TDMA, FDMA, etc.)
Part of the theoretical and intellectual appeal, admittedly, is also the fact that performance of achievable schemes and converses simplify as $\kappa\rightarrow 0$.

Now, it should be clear that at $\kappa=0$, there could be no communication, and as a result the squared error distortion is exactly $1$ (assuming a unit variance source). Therefore, using a first order approximation with respect to $\kappa$, we expect the distortion to behave as 
\[
D_\kappa(N) \approx 1+ \kappa \left. \dv{D_\kappa(N)}{\kappa}\right|_{(\kappa=0)} 
\]
when $\kappa$ is small but non-zero\footnote{The derivative is always negative as the distortion can only be improved with positive bandwidth.}.
Thus, the quantity of interest throughout the paper will be the {\em fidelity} the coding scheme achieves, defined as the negative slope of the distortion at $\kappa=0$, i.e., 
\[
F(N) = -  \left. \dv{D_\kappa(N)}{\kappa}\right|_{(\kappa=0)}. 
\]
It will also prove more convenient to describe the fidelity as a function of {\em quality} level $Q=\frac{1}{N}$, i.e., as $F(Q)$.
Our goal then is to analyze the minimum power needed to achieve a given {\em fidelity-quality profile} ${\cal F}(Q)$, i.e., to ensure 
\[
F(Q) \geq {\cal F}(Q) \; .
\]

We derive a family of lower bounds to the minimum achievable power for a general profile ${\cal F}(Q)$, and discuss certain profiles in more detail. 
Specifically, we show that (i) the optimal scheme for rational profile with order one is simple uncoded transmission, and (ii) establish upper and lower bounds on the minimum energy for rational profile with order two\footnote{We refer to  ${\cal F}(Q)=\frac{\alpha Q}{1+\alpha Q}$ and ${\cal F}(Q)=\frac{\alpha Q^2}{1+\alpha Q^2}$ for some $\alpha$ as rational profiles with order one and two, respectively.}.

One of the similar universal coding scenarios in the literature is given in \cite{Woyach2012}, where a maximum regret approach for compound channels is proposed. The objective in their scenario is to minimize the maximum ratio of the capacity to the achieved rate at any noise level.
Other related work in the literature includes \cite{erman2015}, \cite{Eswaran2007}, \cite{Misra2012}, \cite{Lomnitz2011}, \cite{erman2017}, \cite{wilson2007}, and \cite{Wang2009}. We will explain some of these  in details in section III.

The rest of the paper is organized as follows. The next section is devoted to preliminaries and notation. In Section~III, we review the related work. In Section~IV, a simple lower bounds on $P_{min}(\cal{F})$ is derived.
Finally, in Section~V, we analyze rational fidelity-quality profiles of order one and two and propose upper and lower bounds for them.

\section{Preliminaries and Notation}

Let $X^n$ be an i.i.d.\ unit-variance Gaussian source to be transmitted over an AWGN channel $V^m=U^m+W^m$, where $U^m$ is the channel input, $W^m\sim\mathcal{N}\left(\mathbf{0},N\mathbf{I}_m\right)$ is the additive noise, and $V^m$ is the observation at the receiver. 
\begin{defn}
\label{defn:Achievability}
A pair of distortion-noise profile ${\cal D}_\kappa(N)$ and power level $P$ is said to be {\em achievable} if for every $\epsilon>0$, there exists 
$(m,n)$, an encoder
\[
f^{m,n}: \mathbb{R}^n\longrightarrow \mathbb{R}^m \; ,
\]
and decoders
\[
g^{m,n}_{N}: \mathbb{R}^m\longrightarrow \mathbb{R}^n
\]
for every $0<N<\infty$, such that
\[
\frac{m}{n} \leq \kappa+\epsilon
\]
together with
\[
\frac{1}{m} \mathbb{E}\left\{||f^{m,n}(X^n)||^2\right\} \leq P+\epsilon
\]
and
\[
\frac{1}{n}\mathbb{E}\left\{||X^n-g^{m,n}_{N}(f^{m,n}(X^n)+W^m_{N})||^2\right\} \leq {\cal D}_\kappa(N)+\epsilon
\]
with $W^m_{N}$ being the i.i.d.\ channel noise with variance $N$. 
\end{defn}

For a given function ${\cal D}_\kappa$, the main quantity of interest would be
\[
P_{\min}({\cal D}_\kappa) = \inf \{P : ({\cal D}_\kappa,P) \mbox{ achievable}\} 
\]
with the understanding that $P_{\min}({\cal D}_\kappa) =\infty$ if there is no finite $P$ for which $({\cal D}_\kappa,P)$ is achievable.

As mentioned in the Introduction, we investigate this problem at the extreme of $\kappa\rightarrow 0$, in which case no distortion level less than $1$ can be achieved for any $N$, and the problem as it is stated becomes trivialized. 
We instead look into what can be achieved for near-zero $\kappa$ in terms of {\em how fast the distortion decreases} as a function of $\kappa$ for all levels of noise $N>0$, or equivalently, for all levels of quality $Q=\frac{1}{N}$.
\begin{defn}
\label{defn:Achievability2}
A pair of fidelity-quality profile ${\cal F}(Q)$ and power level $P$ is said to be {\em achievable} if for every $\epsilon>0$, there exists an achievable $({\cal D}_{\kappa},P)$ for all $0\leq\kappa<\epsilon$ such that 
\[
{\cal F}(Q) =  -  \left. \dv{{\cal D}_\kappa(\frac{1}{Q})}{\kappa}\right|_{(\kappa=0)} \; .
\]
\end{defn}

Note that ${\cal D}_{\kappa}(N)$ needs to be differentiable at $\kappa=0$. Also, $P_{\min}({\cal F})$ is similarly defined as 
\[
P_{\min}({\cal F}) = \inf \{P : ({\cal F},P) \mbox{ achievable}\} \; .
\]

\section{Related Work}
In this section, we review the previous work.

In \cite{erman2015}, the tradeoff between the distortion when the channel quality is good versus bad is investigated for transmission of memoryless Gaussian sources over channels with additive white Gaussian noise (AWGN). They propose novel schemes for $\frac{1}{2} \leq \kappa \leq 1$ and $1 \leq \kappa \leq 2$ achieving nontrivial tradeoffs outperforming all known schemes. 

In \cite{erman2017}, lossy transmission of a memoryless bivariate Gaussian source over a bandwidth mismatched AWGN  channel with two receivers is studied. The authors show that their scheme for bandwidth compression outperforms the HDA coding scheme of \cite{Behroozi2011} if their proposed conjecture (supported by numerical observations) is indeed true.

In \cite{wilson2007}, the problem of broadcasting a Gaussian source to two users over an AWGN channel is considered. A framework is developed which shows a duality between source-channel coding schemes for bandwidth expansion ($\kappa > 1$) and those for bandwidth compression ($\kappa <1$). The authors then utilized the bandwitdh expansion scheme proposed by Reznic, Zamir, and Feder in \cite{Reznic2006} to develop achievable schemes for $\kappa<1$. The authors also provide an analysis of performance of source-channel coding schemes in the presence of a signal-to-noise ratio (SNR) mismatch.

In \cite{Wang2009}, three hybrid digital-analog (HDA) systems for the transmission of Gaussian sources over AWGN channels  under bandwidth compression are studied. Upper bounds on the asymptotically optimal mean squared error distortion are calculated for both matched and mismatched channel conditions. 

As we discussed previously, we target robust communication with noise variance $N$ is unknown at the transmitter. At a first glance, it seems that we can employ the results in the aforementioned work to obtain achievability results simply by letting $\kappa \rightarrow 0$. However, they all are limited to those values of $N$ for which at least one digital layer would be decoded (i.e., $N\leq N_1$ for some $N_1$), and sacrifice the distortion when $N>N_1$ . We instead need a scheme which can operate at {\em all} noise levels $N>0$. Towards that end, we develop our own achievability schemes. 

\section{A Lower Bound on $P_{\min}({\cal F})$}
\label{sctn:LowerBounds}

An immediate lower bound on $P_{\min}({\cal F})$ follows from (\ref{dformula}). Despite its simplicity, it will be instrumental in the sequel.

\begin{lmm}
\begin{equation}
\label{eqtn:pmin1stOrder2}
P_{\min}({\cal F}) \geq \sup_{Q>0} \frac{\exp({\cal F}(Q))-1}{Q}  \; .
\end{equation}
\end{lmm}
\begin{IEEEproof}
From (\ref{dformula}), it follows that for any fixed $N_0$, the distortion $D_\kappa(N_0)$ achieved by any scheme with bandwidth $\kappa$ has to satisfy
\begin{equation}
\label{lowerboundtrivial}
D_\kappa(N_0) \geq  \frac{1}{\left(1+\frac{P}{N_0}\right)^\kappa} \; .
\end{equation}
Defining $Q_0=\frac{1}{N_0}$ and 
\[
F(Q_0)=-\dv{D_\kappa\left(\frac{1}{Q_0}\right)}{\kappa}\bigg|_{(\kappa=0)} \; ,
\]
we can approximate (\ref{lowerboundtrivial}) around $\kappa \approx 0$ as 
\[
1 - \kappa F(Q_0) \geq 1- \kappa \ln \left(1+PQ_0\right) \; .
\]
Since $Q_0>0$ is arbitrary, this implies that $({\cal F},P)$ is achievable only if
\[
{\cal F}(Q) \leq \ln (1+PQ)
\]
for all $Q>0$. The result (\ref{eqtn:pmin1stOrder2}) then follows by rearranging.

\end{IEEEproof}

\section{Analysis for Specific Profiles}
\subsection{Rational Profile with Order One}

Consider the fidelity-quality profile given as
\begin{equation}
{\cal F}(Q)=\frac{\alpha Q}{1+\alpha Q} \; .
\label{profileq}
\end{equation}
In what follows we show that a simple uncoded transmission in fact achieves $P_{\min}({\cal F})$, and therefore is optimal. 

\begin{lmm}
\label{lmma:Linear}
$P_{\min}({\cal F}) = \alpha$ for the profile given in (\ref{profileq}). Moreover, uncoded transmission with $m=1$ and 
\begin{align}
\label{uncoded}
U=\sqrt{\frac{\alpha}{n}} \sum_{t=1}^{n} X_t
\end{align}
achieves the minimum power.
\end{lmm}
\begin{IEEEproof}
Clearly, uncoded transmission as described in (\ref{uncoded}) uses a bandwidth factor of $\kappa=\frac{1}{n}$, and expends power $\alpha$. It can easily be shown that the resultant expected distortion given by
\[
D_\kappa(N)=1-\kappa \frac{\alpha}{\alpha+N}
\]
for any $\kappa\geq 0$ and $N>0$, translating into 
\[
F(Q)  = \left.-\dv{D_\kappa(\frac{1}{Q})}{\kappa}\right|_{(\kappa=0)}=\frac{\alpha Q}{1+ \alpha Q}
\]
for all $0<Q<\infty$.
Since this coincides (and hence complies) with ${\cal F}(Q)$, we conclude that $P_{\min}({\cal F})\leq\alpha$.

To show that $P_{\min}({\cal F})\geq\alpha$, it suffices to use the lower bound (\ref{eqtn:pmin1stOrder2}):
\begin{eqnarray*}
P_{\min}({\cal F}) & \geq & \sup_{Q>0} \frac{\exp({\cal F}(Q))-1}{Q} \\
& = & \sup_{Q>0} \frac{\exp(\frac{\alpha Q}{1+ \alpha Q})-1}{Q}  \\
& \geq & \lim_{Q\rightarrow 0}\frac{\exp(\frac{\alpha Q}{1+ \alpha Q})-1}{Q}  \\
& = & \lim_{Q\rightarrow 0}\frac{\alpha \exp(\frac{\alpha Q}{1+ \alpha Q})}{(1+\alpha Q)^2}  \\
\\
& = & \alpha \; .
\end{eqnarray*}
\end{IEEEproof}

Lemma~\ref{lmma:Linear} may not be surprising as the profile ${\cal F}(Q)$ in (\ref{profileq}) is ``tailored'' to the performance of uncoded transmission.
Nevertheless, just as in the energy-distortion context in~\cite{ermanenergy}, it is a somewhat surprising example where uncoded transmission is optimal in any context other than {\em matched bandwidth} scenarios.

\subsection{ Rational Profile with Order Two}
In this section, we consider the fidelity-quality profile given as
\begin{equation}
{\cal F}(Q)=\frac{\alpha Q^2}{1+\alpha Q^2} \; .
\label{profileq2}
\end{equation}

We begin by lower bounding $P_{min}(\cal{F})$.
According to (\ref{eqtn:pmin1stOrder2}), we can write
\begin{equation}
\label{lowernumerical}
P_{\min}({\cal F}) \geq \sup_{Q>0} \frac{\exp(\frac{\alpha Q^2}{1+\alpha Q^2})-1}{Q}  \; .
\end{equation}
Unfortunately, it is not easy to solve the optimization problem in (\ref{lowernumerical}) analytically. Therefore, we solve it numerically and plot it as a function of $\alpha$ in Figure 1.
 
Towards developing an achievable scheme for the profile $P_{\min}(\cal{F})$, and therefore obtaining an upper bound for it, we first show that it is not possible to achieve  $P_{\min}(\cal{F})$ using purely uncoded transmission.
That is because uncoded transmission with power $P$, as analyzed in Lemma~\ref{lmma:Linear}, would achieve a fidelity of $\frac{PQ}{1+PQ}$.  Thus, $P$ needs to satisfy
\begin{equation}
\label{onlyanalog}
    \frac{PQ}{1+PQ} \geq \frac{\alpha Q^2}{1+\alpha Q^2} 
\end{equation}
for all $Q>0$, which simplifies to $P \geq \alpha Q$. But this is not possible with a finite $P$.

To remedy this, we propose a hybrid scheme with one digital and one analog transmission layer.
We describe it for any fixed $\kappa=\frac{m}{n}$, but then specialize its performance to $m=1$ and $n\rightarrow\infty$, and thus to $\kappa\rightarrow 0$.
We divide the available power as $P=P_a + P_1$, where $P_a$ and $P_1$ are the power levels of the analog and digital layers, respectively. 
For any $(m,n)$, we treat $X^n$ and $U^m$ as super symbols in our mapping of long source blocks of length $nl$ onto channel words of length $ml$, where $l$ is large enough to approach the Shannon limits. 
We quantize $X^{nl}$ using to the super-letter distribution 
\begin{equation}
\label{source}
X^n=S_1^n + E_1^n \nonumber
\end{equation}
with $S_1^n \perp E_1^n $. Note that this constrains the covariance matrices of $E_1^n$ and $X^n$ such that
\begin{equation}
\label{fffff}
\mathbf{0} \leq \mathbf{C}_{E_1^n} \leq \mathbf{C}_{X^n}. \nonumber
\end{equation}
We assume that $\mathbf{C}_{X^n}=\mathbf{I}$ which gives us the following constraint,
\begin{equation}
\label{bbbbb}
\mathbf{0} \leq \mathbf{C}_{E_1^n} \leq \mathbf{I}. 
\end{equation}
Now, we use an $m \times n$ matrix $\mathbf{K}$ to transmit $X^n $ using
\begin{equation}
\label{ccccc}
U_a^m= \mathbf{K} X^n\nonumber
\end{equation}
such that 
\begin{equation}
\label{ddddd}
\frac{1}{m} E[\norm{U_a^m}^2]= \frac{1}{m}{\rm Tr}(\mathbf{K} \mathbf{K}^T)=P_a\nonumber
\end{equation}
Let $U_1^{ml}$ denote the codeword for conveying $S_1^{nl}$ such that 
\begin{equation}
\label{eeeee}
\frac{1}{m} E[\norm{U_1^m}^2]= P_1 .\nonumber
\end{equation}
This codeword is superimposed on $U_a^{ml}$ using dirty-paper coding where $U_a^{ml}$ is treated as channel state information (CSI) known at the encoder.

Let $V^{ml}=U_a^{ml} + U_1^{ml} + W_N^{ml}$ be the received vector. 
We designate a noise threshold $N_1$ such that if  $N \leq N_1$, the digital information (i.e., the quantized block $S_1^{nl}$) is successfully decoded, and otherwise reconstruction should rely purely on analog information. 

Thus, for $N > N_1$, the MMSE estimator is given by
\begin{align}
    \hat{X}^n=\mathbf{A}_1 V^m = \mathbf{A}_1(U_a^m + U_1^m + W_N^m)\nonumber,
\end{align}
where 
\begin{align}
   & \mathbf{A}_1=\mathbf{C}_{X^n V^m} \mathbf{C}_{V^m}^{-1} \nonumber \\
    &\mathbf{C}_{X^n V^m}=\mathbf{K}^T \nonumber \\
    &\mathbf{C}_{V^m}^{-1} =\big(\mathbf{K}\mathbf{K}^T+(P_1+N)\mathbf{I}\big)^{-1}.
\end{align}
The corresponding distortion is given by
\begin{align}
\label{distortion1}
D(N)&=\frac{1}{n} \sum_{t=1}^{n} E[(X_t-\hat{X_t})^2]\nonumber \\
&=\frac{1}{n}\big[{\rm Tr}(\mathbf{C}_{X^n})-{\rm Tr}(\mathbf{A}_1 \mathbf{K}\mathbf{C}_{X^n})\big]\nonumber \\
&=\frac{1}{n}\bigg[n-{\rm Tr}\big(\mathbf{K}^T(\mathbf{K}\mathbf{K}^T+(P_1+N)\mathbf{I})^{-1}\mathbf{K}\big)\bigg].
\end{align}

For $N \leq N_1$, on the other hand, one can conclude using standard arguments in dirty-paper and Wyner-Ziv coding that to be able to transmit $S_1^{nl}$ successfully to the receiver, we need
\begin{eqnarray}
\lefteqn{\frac{m}{2n} \log(1+\frac{P_1}{N_1})} \;\;\;\;\; \nonumber \\
&\geq& \frac{1}{n} I(X^n;S_1^n|V^m) \nonumber \\
&=&\frac{1}{n} [I(X^n;S_1^n)-I(V^m;S_1^n)] \nonumber \\
&=&\frac{1}{n}[h(X^n)-h(E_1^n)-h(V^m)+h(V^m|S_1^n)] \nonumber \\
&=&\frac{1}{2n}\log \frac{\det(\mathbf{C}_X^n) \det (\mathbf{K}\mathbf{C}_{E_1}^n \mathbf{K}^T + (P_1+N_1)\mathbf{I})}{\det(\mathbf{C}_{E_1}^n) \det (\mathbf{K}\mathbf{C}_{X}^n \mathbf{K}^T + (P_1+N_1)\mathbf{I})} \nonumber \\
\label{constraint1}
&=&\frac{1}{2n}\log \frac{ \det (\mathbf{K}\mathbf{C}_{E_1}^n \mathbf{K}^T + (P_1+N_1)\mathbf{I})}{\det(\mathbf{C}_{E_1}^n) \det (\mathbf{K} \mathbf{K}^T + (P_1+N_1)\mathbf{I})}. 
\end{eqnarray}

The resultant MMSE estimator is given by
\begin{align}
    \hat{E}_1^n=\mathbf{A}_2 \Tilde{V}^m = \mathbf{A}_2(U_1^m +\mathbf{K} E_1^n+ W_N^m)\nonumber,
\end{align}
where 
\begin{align}
   & \mathbf{A}_2=\mathbf{C}_{{E^n_1} \Tilde{V}^m} \mathbf{C}_{\Tilde{V}^m}^{-1} \nonumber \\
    &\mathbf{C}_{{E^n_1} \Tilde{V}^m}= \mathbf{C}_{E^n_1} \mathbf{K}^T \nonumber \\
    &\mathbf{C}_{V^m}^{-1} =\big(\mathbf{K} \mathbf{C}_{E^n_1} \mathbf{K}^T+(P_1+N)\mathbf{I}\big)^{-1}.
\end{align}
The corresponding distortion is given by
\begin{eqnarray}
\lefteqn{D(N)} \\ &=& \frac{1}{n} \sum_{t=1}^{n} E[(E_{1t}-\hat{E_{1t}})^2]\nonumber \\
&=& \frac{1}{n}\big[{\rm Tr}(\mathbf{C}_{E_1^n})-{\rm Tr}(\mathbf{A}_2 \mathbf{K}\mathbf{C}_{{E_1}^n})\big]\nonumber \\
&=& \frac{1}{n}\bigg[{\rm Tr}(\mathbf{C}_{E_1^n})\nonumber \\
&& -{\rm Tr}\big(\mathbf{C}_{E_1^n} \mathbf{K}^T(\mathbf{K} \mathbf{C}_{E_1^n} \mathbf{K}^T+(P_1+N)\mathbf{I})^{-1}\mathbf{K}\mathbf{C}_{E_1^n}\big)\bigg] \; .\nonumber \\
\label{distortion2}
\end{eqnarray}

The scheme described up until this point is general enough to be used for any bandwidth factor $\kappa$ (in fact, even for bandwidth expansion).
The next theorem states achievable fidelity levels as a function of quality $Q$ when $\kappa\rightarrow 0$.

\begin{thm}
The pair $({\cal F},P)$ is achievable if there exists a triplet $(P_a,P_1,Q_1)$ such that $P=P_a+P_1$ and $F(Q)\geq{\cal F}(Q)$ where 
\[
F(Q) = \left\{\begin{array}{ll}
\frac{P_a Q}{1+P Q} & 0<Q<Q_1 \\
\ln(1+P_1 Q_1) + \frac{P_a Q}{1+P Q} & Q\geq Q_1
\end{array}\right. \; .
\]
\end{thm}

\begin{IEEEproof}
Let $m=1$ and therefore $\kappa=\frac{1}{n}$ in the scheme described above. 
We also choose 
\[
\mathbf{K} = \left[\begin{array}{cccc} k & k & \cdots & k\end{array} \right]
\]
and  $\mathbf{C}^n_{E_1}=\sigma^2_1 \mathbf{I}_n$ with some $\sigma^2_1\leq 1$ to be specified below. 

The distortion in (\ref{distortion1}) then simplifies to 
\[
D(Q) =1-\kappa \frac{P_a Q}{1+P Q} 
\]
for all $0<Q<Q_1$, resulting in 
\begin{equation}
\label{fidelity1}
F(Q) = \left.-\frac{dD(Q)}{d\kappa}\right|_{\kappa=0}=\frac{P_a Q}{1+P Q} \; .
\end{equation}

On the other hand, for $Q \geq Q_1$, (\ref{distortion2}) and (\ref{constraint1}) reduce to 
\begin{align}
\label{distortion22}
D(Q)&={\sigma}^2_1 (1-\kappa \frac{\sigma^2_1 P_a Q}{1+(\sigma^2_1 P_a +P_1)Q}),
\end{align}
and
\begin{align}
\label{C1}
1+P_1 Q_1 \geq \frac{1+(\sigma^2_1 P_a +P_1)Q_1}{(\sigma^2_1)^n(1+P Q_1)}
\end{align}
respectively, where $n=\frac{1}{\kappa}$.
It is straightforward to show that $D(Q)$ in (\ref{distortion22}) is increasing in $\sigma^2_1$. Thus, the minimum $D(Q)$ is achieved by minimum $\sigma^2_1$ satisfying (\ref{C1}). 
For convenience, let $\beta=\sigma^2_1$ and rewrite (\ref{C1}) as
\begin{equation}
    f_n(\beta)=\beta^n(1+P Q_1)(1+P_1 Q_1)-\beta P_a Q_1-(1+ P_1 Q_1)\geq 0.\nonumber
\end{equation}
Since $f_n(0)<0$ and $f_n(1)>0$, there has to be a $0 < \beta^*_n < 1$ such that $f_n(\beta^*_n)=0$. Furthermore, it is easy to show that $\beta^*_n$ is unique and $\beta^*_n \rightarrow 1$ as $n \rightarrow \infty$. Therefore, we can approximate $\beta^*_n$ as the solution to
\begin{equation}
    \hat{f}_n(\beta)=\beta^n(1+P Q_1)(1+P_1 Q_1)-P_a Q_1-(1+ P_1 Q_1)= 0\nonumber
\end{equation}
instead. It can therefore be seen that $\beta^*_n \approx a^{\frac{1}{n}}$ for very large $n$, where
\[
    a = \frac{1}{1+P_1 Q_1} \; .
\]
In fact, simply choosing $\beta_n=a^{\frac{1}{n}}$ readily satisfies $f_n(\beta_n)\geq 0$ for all $n$, so we do not need to make this approximation more precise mathematically. 

We can then rewrite (\ref{distortion22}) as a function of $\kappa$ as 
\[
D_\kappa(Q) = a^{\kappa}\bigg(1-\kappa \frac{a^{\kappa}P_a Q}{1+(a^{\kappa}P_a+P_1)Q}\bigg)\nonumber
\]
for small $\kappa$. 
Hence,
\begin{equation}
\label{fidelity2}
F(Q) = \left.-\frac{dD_\kappa(Q)}{d\kappa}\right|_{\kappa=0} =\ln(1+P_1 Q_1) + \frac{P_a Q}{1+P Q}.
\end{equation}
Bringing together (\ref{fidelity1}) and (\ref{fidelity2}) finishes the proof.
\end{IEEEproof}

We would like to draw the parallel to the achievable scheme presented in \cite{ermanenergy} in the context of energy-distortion tradeoff. 
In the achievable scheme we presented, there is a persistent behavior of $\frac{P_a Q}{1+P Q}$, very much like the piecewise linear behavior in the energy-distortion case. This behavior is disrupted by a ``jump'' of magnitude $\ln(1+P_1 Q_1)$ after $Q=Q_1$, also as in the energy-distortion case.

Figure 2 depicts how the proposed achievable scheme can {\em potentially} comply with the rational profile ${\cal F}(Q)$ of order two, with properly chosen $(P_a,P_1,Q_1)$.
We next show this can indeed be done. 
For $0 \leq Q < Q_1$, we need
\[
 \frac{P_a Q}{1+P Q} \geq  \frac{\alpha Q^2}{1+\alpha Q^2} 
\]
which can be simplified to  
\[
 P_a \geq  \alpha Q+ \alpha P_1 Q^2 \; .
\]
Since the right-hand side is increasing in $Q$, it suffices to have 
\begin{equation}
\label{cc1}
     P_a \geq \alpha Q_1+ \alpha Q_1^2 P_1 \; .
\end{equation}

On the other hand, for $Q \geq Q_1$, we must satisfy
\[
   \ln(1+P_1 Q_1) + \frac{P_a Q}{1+P Q} \geq  \frac{\alpha Q^2}{1+\alpha Q^2} 
\]
or rearranging, $P_a \geq g(Q)$ with
\[
g(Q)=\frac{\big[\frac{\alpha Q^2}{1+\alpha Q^2}-\ln({1+P_1Q_1})\big](1+P_1Q)}{Q \big[\frac{1}{1+\alpha Q^2}+\ln({1+P_1Q_1})\big]}.
\]

Since finding the maximum of $g(Q)$ over $Q \geq Q_1$ analytically seems difficult, we upper bound it as $z(Q) \geq g(Q)$ where 
\begin{align}
    z(Q)&=\frac{\alpha Q^2 (1+P_1 Q)}{\alpha Q^3 \ln{(1+P_1 Q_1)}} \nonumber \\
    &=\frac{1}{\ln{(1+P_1 Q_1)}}\bigg(\frac{1}{Q}+P_1\bigg)
\end{align}
Since $z(Q)$ is decreasing for $Q \geq Q_1$, $P_a\geq z(Q)$ is the same as 
\begin{align}
\label{c22}
    P_a \geq \frac{1}{\ln{(1+P_1 Q_1)}}\bigg(\frac{1}{Q_1}+P_1\bigg)
\end{align}
Combining (\ref{cc1}) and (\ref{c22}), we obtain
\begin{align}
    P_a \geq \max \bigg[\frac{1}{\ln{(1+P_1 Q_1)}}\bigg(\frac{1}{Q_1}+P_1\bigg),\alpha Q_1+ \alpha Q_1^2 P_1\bigg]
\end{align}
as the minimum possible $P_a$ for any choice of $(P_1,Q_1)$. Since this maximum is finite, it is indeed possible to achieve ${\cal F}(Q)$ using the proposed hybrid scheme. 

By searching through the space of $P_1,Q_1>0$, we obtained the minimum possible $P=P_a+P_1$ as a function of $\alpha$, which is depicted in Figure 1.
The gap between the lower and upper bounds appear to saturate to a constant around 13dB.

\subsection{Extension of the Achievable Scheme to $K$ Layers}

\begin{figure}
  \centering
    \includegraphics[width=0.52\textwidth]{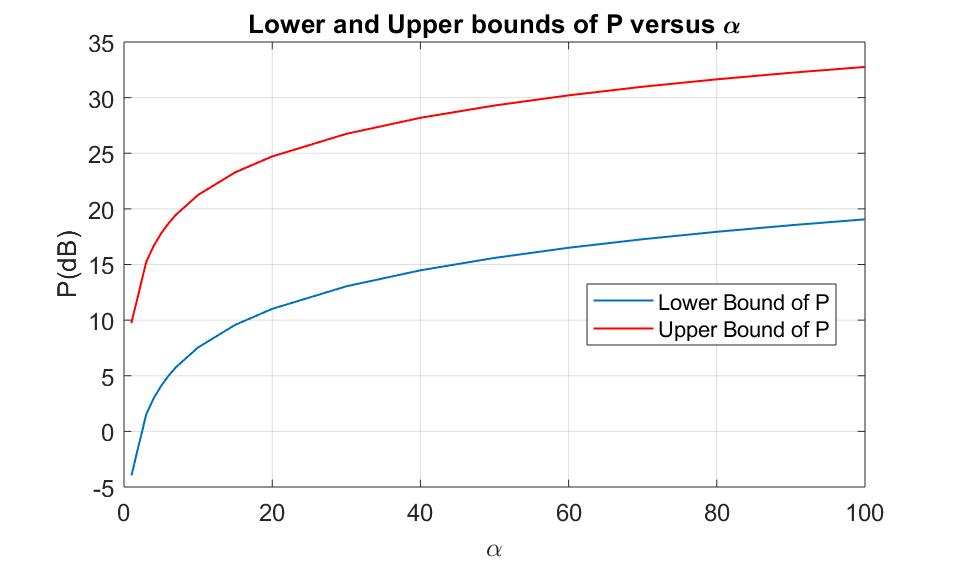}
    \caption{Lower and upper bounds to $P$ (in dB) as a function of $\alpha$.}
\end{figure}

\begin{figure}
  \centering
    \includegraphics[scale=0.45]{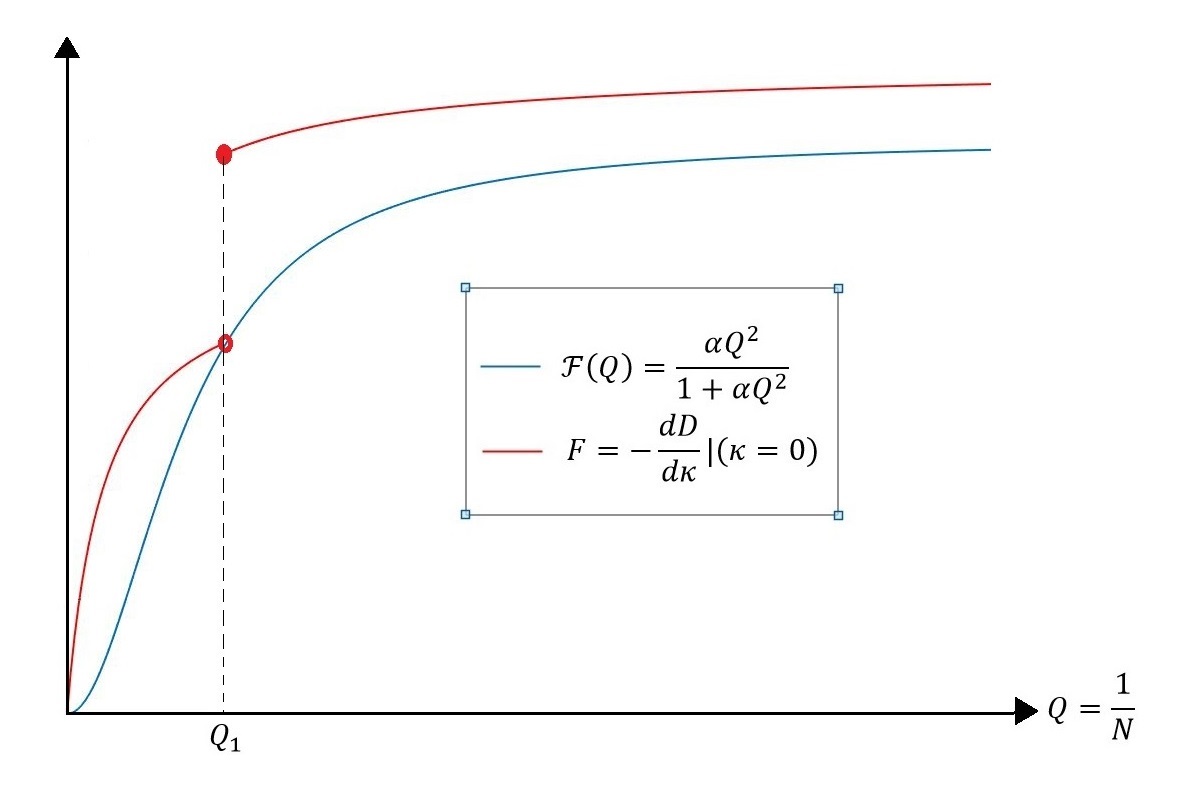}
    \caption{The achieved fidelity $F(Q)$ versus  ${\cal F}(Q)=\frac{\alpha Q^2}{1+\alpha Q^2}$.}
\end{figure}

The proposed hybrid scheme can be extended into multiple layers of digital information by simply quantizing the quantization error from the previous round and building a coding hierarchy where the $k$th layer ``sees'' the channel words of the layers below it as noise and above it as interference that can be canceled by virtue of dirty paper coding. 
Each layer $k$ will bring about a similar jump in the fidelity at some quality level $Q_k$.
Due to lack of space,  we only provide the corresponding $F(Q)$.
For $0<Q<Q_1$,
\begin{align}
F(Q) = d_0 \stackrel{\Delta}{=} \frac{P_a Q}{1+P Q}.\nonumber 
\end{align}
For $Q_k \leq Q < Q_{k+1}$, and $ 1 \leq k< K$,
\begin{align}
F(Q) = d_k \stackrel{\Delta}{=} d_{k-1}+\ln \bigg(\frac{1+(\sum_{i=k}^{K} P_i)Q_k}{1+(\sum_{i=k+1}^{K} P_i)Q_k}\bigg)\nonumber
\end{align}
Finally, for $Q\geq Q_K$,
\begin{align}
F(Q) = d_K \stackrel{\Delta}{=} d_{K-1}+\ln \bigg(1+P_K Q_K\bigg).
\end{align}

\newpage

\end{document}